\documentstyle[aps]{revtex}
\advance\hsize by 0.0truecm  \hoffset=0.7cm
\begin{document}
\bibliographystyle{prsty}
\preprint{\small  $ \begin{array}{r} \rm  RUB-TPII-08-95
         \\  \rm  SUNY-NTG-95-28
         \\  \rm hep-ph/9605316   \end{array}   $
           }
\title{\large A Top Quark Soliton and
\\ its Anomalous Chromomagnetic Moment}
\author{
Jochen Berger$^{(1)}$,
\footnote{email:jochenb@hadron.tp2.ruhr-uni-bochum.de},
Andree Blotz $^{(2)}$,
\footnote{email:andreeb@luigi.physics.sunysb.edu},\\
Hyun-Chul Kim$^{(1)}$,
\footnote{email:kim@photon.tp2.ruhr-uni-bochum.de}
and Klaus Goeke$^{(1)}$
\footnote{email:goeke@hadron.tp2.ruhr-uni-bochum.de}
}
\address{(1)
Institute for Theoretical  Physics  II,   P.O. Box 102148,  \\
Ruhr-University Bochum,
 D-44780 Bochum,   GERMANY
       }
\address{(2)
Department of Physics, State University of New York,\\ Stony Brook,
New York 11794,   U.S.A.
       }
\date{\today}
\maketitle
\begin{abstract}
{
We show that under the assumption of dynamical
symmetry breaking of electro weak interactions
by a top quark
condensate, motivated by the Top Mode Standard Model, the top
quark in this effective theory can be considered then as
chiral color soliton (qualiton). This is
realized in an effective four-fermion
interaction with chiral $SU(3)_c$  as well as
$SU(2)_L\otimes U_Y(1)$ symmetry. In the pure top sector
the qualiton
consists of a top valence
quark and a Dirac sea of top and anti-top quark coupled to
a color octet of Goldstone pions.
The mass spectra, isoscalar quadratic radii and the
anomalous chromomagnetic moment due to a non-trivial
color form factor are calculated with 
zero and finite current top masses
and effects at the Hadron Colliders are discussed. The anomalous
chromomagnetic moment turns out to have a value consistent with the
top production rates of the D0- and CDF-measurements.}
\end{abstract}
%
%
\section{Introduction }
That the mechanism for
symmetry breakdown in the electro weak theory of the standard model 
(SM) is maybe of dynamical origin was first noted by
Nambu \cite{bolo:nambu} and Miransky et al. \cite{bolo:miransky}
and is presently under intensive discussion
\cite{bolo:zhang1,bolo:bhl,bolo:clro,bolo:liro,bolo:pepezh}.
The clue is that the recently measured top quark mass of
CDF Collaboration \cite{bolo:cdf2}
$m_t=176\pm 8\pm 10{\rm GeV}$ ($\sigma_p=6.8^{+3.6}_{-2.4}pb$) 
and D0
Collaboration \cite{bolo:d0} $m_t=199^{+19}_{-21}\pm
22{\rm GeV}$ ($\sigma_p=6.4\pm 2.2pb$) turn out to be correct and 
the top mass
 is of the order of the Fermi scale
$v=\left(\sqrt{2}G_F\right)^{-1/2}=246{\rm GeV}$,  the scale of
electro weak symmetry breaking.
Therefore if there is any new physics in the electro weak theory
beyond this mass scale, the top quark should be  an ideal probe for
theses effects   which are almost invisible for the lighter
quarks.

Also, up to now there is no evidence 
for any process with Higgs boson
contribution at Born level \cite{bolo:schaile}.
Furthermore if the D0 Measurement turns out to be correct,
the bounds on the Higgs will be in 
the TeV regime \cite{bolo:schaile},
whereas baryon number violating processes 
in the minimal Standard model
(cf. Diakonov et al. \cite{bolo:dss}) presumably restrict
$m_{H}$ to be smaller than $80{\rm GeV}$. Having this in mind it 
is not
too  speculative to think of new physics 
beyond the electroweak theory.

For this purpose we want to consider a model Lagrangian on
the electroweak  scale  hoping that it 
contains the relevant degrees of
freedom of the real unified theory at some higher energy scale. 
The low energy
behaviour should of course coincide with the standard electro weak
theory \cite{bolo:yuan}. One of the 
candidates for such an effective 
theory is the so called {\it Top Mode Standard Model}
\cite{bolo:nambu,bolo:miransky}
where the dynamical symmetry breaking is performed via a top quark 
condensate in a BCS oder Nambu--Jona-Lasinio 
like theory \cite{bolo:njl1}.
Because the top quark
Yukawa coupling is expected to be of the order of one,
the situation is similar to the
nucleon constituent quarks in QCD.
This makes the assumption of a dynamical symmetry breaking
reasonable, which gives a mass to the top quark, comparable to the
up- and down-constituent quarks in QCD, which are known to get their
masses dynamically.

Recently bounds on the radii for the
up, down, strange, charm and bottom quarks have been reported
\cite{bolo:kopp}  and they are actually very 
small, i.e. $r<10^{-5}fm$.
So they can be considered as point like with no
non-trivial formfactor and therefore vanishing anomalous magnetic
moment. For the top quark the situation is 
probably different due to its
large mass which is very much larger than 
the masses of the other fermions.

It was noted by Brodsky and Drell \cite{bolo:brodre}
and recently by Brodsky and Schlumpf  \cite{bolo:brosch}
that the anomalous magnetic moment is presumably linear in the size
and mass of the composite particle.
If such an picture also applies to the top quark, 
the large mass in conjunction
with a finite radius ($\simeq 10^{-3}{\rm fm}$)  should 
give a measurable effect. 

But actually it was proposed already that
a discrepancy between SM cross section for 
top production \cite{bolo:smith}
and first CDF measurements \cite{bolo:cdf}
can be explained by a non-trivial formfactor for the top
(see Ref. \cite{bolo:dicus2} for a short review).
Therefore we will investigate  here the idea 
proposed by Atwood, Kagan
and Rizzo \cite{bolo:akr} that already a
small anomalous chromomagnetic moment for 
the top quark can enhance the
${\bar t}t$ production cross section
\footnote{See also e.g. \cite{bolo:zhanglee,bolo:whiszhang}
for a discussion of non-standard top-quark couplings.}. We will find,
that our model gives an anomalous magnetic moment, which has the
correct magnitude to explain the production rate at the Tevatron.

The formalism that we will use to calculate
the anomalous chromomagnetic moment for the top quark
is an effective Lagrangian with BCS or Nambu--Jona-Lasinio (NJL)
type mechanism for dynamical symmetry breaking. The idea is that
at some high energy scale $\Lambda$  the 
SM contains only the usual quark,
lepton and gauge boson degrees of freedom, but no fundamental Higgs
scalars \cite{bolo:bhl}.
The dynamical symmetry breaking in the electroweak sector
is then triggered  by the condensate
of  the top.

The formalism of solitons having non-integer baryon number was
suggested by Kaplan
\cite{bolo:kaplan} and later investigated by Gomelski and Karliner
\cite{bolo:gom} and Keaton \cite{bolo:keaton}
in the framework of the {\it Skyrme model},
where the baryon number
is induced topologically via the winding 
number of the chiral profile.
The constituent quarks obtained their masses
via spontaneously broken chiral SU(3) color symmetry.
In the early work \cite{bolo:kaplan} Kaplan wants  to calculate the
properties of the
three lightest constituent quarks in order to deduce finally
the properties of the hyperons which should consist of these
constituent quarks. However, apart from the problem of 
having too large constituent
quark masses after the quantization, the problem of treating the
baryons as bound states of these constituent 
quarks was never solved.

Recently the formalism
of fractional baryon number was refreshed by Zhang
\cite{bolo:zhang1} for the top quark in the electro weak theory.

In the  Top Mode Standard Model (TMSM) (see e.g.
\cite{bolo:bhl,bolo:clro,bolo:liro,bolo:cvetic,bolo:boenic})
 the chiral color symmetry is used to trigger
the formation of a top quark condensate via  a 
local {\it four point fermion}
interaction Lagrangian. However the 
original formulation suffered from
some fine-tuning problems and an enormous ultraviolet cutoff.

We want to make these ideas more explicit by considering the
fermion determinant of an effective 
top quark interaction Lagrangian,
suggested by Lindner and Ross \cite{bolo:liro},
and look for selfconsistent solitonic solutions of the classical
Euler Lagrange equations.
In contrast to the treatment of this Lagrangian 
in the Top Mode Standard
Model, where a top quark condensate leads to 
symmetry breaking and the
top mass is given by a effective potential or the so called gap
equation, we obtain in our approach a localized 
solution of the corresponding solitonic 
equations of motion, the top quark
soliton (qualiton). The top quark constituent mass will be -- as we
shall see later -- the free parameter of our model.
The picture that emerge is that of a
top valence quark carrying around with it a cloud of
Goldstone bosons interacting with the polarized Dirac sea of
top and antitop quarks. The fractional
baryon number $1/3$ of this soliton is then non-topological 
and induced by the
explicit occupation of the discrete valence level.

Adopting the semiclassical quantization 
scheme \cite{bolo:anw} in order to
project out of the soliton a state with definite 
color and spin, we calculate the
mass of the top quark in the fundamental representation and the
low lying resonances.
Furthermore we look at observables like the isoscalar
quadratic radius and the chromomagnetic moment.
The latter quantity is of special interest since it should affect
top production rates and distributions at Hadron Colliders
\cite{bolo:aas,bolo:hp}.

The organization of the paper is as follows. In sect. II we
determine the parameters of our Lagrangian by fixing decay constant
of the color octet of Goldstone bosons. In sect. III we derive the
classical and quantum mass of the top 
states. In sect IV we calculate
the anomalous magnetic moment of the top ground state and in sect. V
we review the numerical results. In sec. VI we finally summarize
our results. The appendix A is devoted to a careful
definition of the quantization of fractional baryon numbers.
\section{The Lagrangian}
In order to account for the physics beyond the Standard Model
several effective Lagrangians 
\cite{bolo:nambu,bolo:miransky,bolo:bhl} 
have been proposed all having in common the
spontaneous breaking of electro weak symmetries by a top quark
condensate. Since we are in this first step 
neither interested in the various gauge
fields itselves nor in the detailed mechanism of spontaneous
breaking of $SU(2)_L\otimes U_Y(1)$
our starting point is,
according to Lindner and Ross \cite{bolo:liro},
the effective interaction (at momenta below a few TeV)
\begin{equation}  
{\cal L}_{int} = - G\ \ {\bar L}_{\alpha,a} \delta^c_{\alpha\beta}
      \gamma^\mu_{ab} L_{\beta,b} \ \
     {\bar t}_{R,{\gamma,c}} \delta^c_{\gamma\delta}
      \gamma^\mu_{cd} t_{R,{\delta,d}}
\label{t1}
\end{equation}
where $L=(t_L,b_L)$ is the lefthanded fermion doublet of the
electroweak theory and greek letters 
imply summation over color indices
and roman letters over spin indices.
As a trivial check, this model has color chiral
$SU(3)_L\otimes SU(3)_R$ symmetry as 
well as $SU(2)_L\otimes U_Y(1)$.
Using Fierz identities such as
\begin{equation}  
   (\gamma^\mu)_{ab}   (\gamma^\mu)_{cd}
    =  (K^\alpha)_{ad}   (K^\alpha)_{cb}, \, 
       K^\alpha=({\bf 1},i\gamma_5,{i\over \sqrt{2}}\gamma_\mu,
              {i\over \sqrt{2}}\gamma_\mu\gamma_5)
\end{equation}
and
\begin{equation}
       \delta^c_{\alpha\beta}
        \delta^c_{\gamma\delta}
         = \frac{1}{2} \sum_{a=0}^8
        (\lambda^a)_{\alpha\delta}
       (\lambda^a)_{\gamma\beta}
\end{equation}
the interaction eq.\ (\ref{t1}) can be rewritten as
\begin{equation}
   {\cal L}_{int} =  G \frac{1}{2}
           \left[ {\bar L} {\bf 1} \lambda^a t
          {\bar t} {\bf 1} \lambda^a L   +
                 {\bar L} i \gamma_5  \lambda^a t
         {\bar t} i \gamma_5 \lambda^a L   \right] .
\end{equation}
Using $\gamma_5 t = t$ and $\gamma_5 L = -L$ this reduces to
\begin{equation}
   {\cal L}_{int} =  G
           \left[ {\bar L} {\bf 1} \lambda^a t
          {\bar t} {\bf 1} \lambda^a L  \right]
\end{equation}
and with the non-chiral decompositions
$L = \frac{1}{2}(1-\gamma_5)Q$ and $t = \frac{1}{2}(1+\gamma_5)q$
we obtain
\begin{eqnarray}
   {\cal L}_{int} &=&  {G \over 4 }
           \left[ {\bar Q}  \lambda^a  q
        +    {\bar Q} \gamma_5  \lambda^a  q   \right]
        \left[
          {\bar q}  \lambda^a Q  -   {\bar q} \gamma_5 \lambda^a Q
       \right]
                 \nonumber \\ &=&
       G  \left[ {\bar Q}  \lambda^a  q
                 {\bar q}   \lambda^a  Q
         -
                {\bar Q} \gamma_5  \lambda^a  q
                {\bar q} \gamma_5  \lambda^a  Q
        +      {\bar Q} \gamma_5  \lambda^a  q
               {\bar q}   \lambda^a  Q
        -      {\bar Q}   \lambda^a  q
               {\bar q} \gamma_5  \lambda^a  Q
           \right] .
\end{eqnarray}
Written in terms of bottom and top fields it is given by
\begin{equation}
  {\cal L}_{int}^{tt}      =
      { G \over 4}  \left[ {\bar t}  \lambda^a  t
                 {\bar t}   \lambda^a  t
         -      {\bar t} \gamma_5  \lambda^a  t
                {\bar t} \gamma_5  \lambda^a  t
                  \right] \qquad \mbox{and}
\end{equation}
\begin{equation}
  {\cal L}_{int}^{tb}      =
      { G \over 4}  \left[ {\bar b}  \lambda^a  t
              {\bar t}   \lambda^a  b
         -     {\bar b} \gamma_5  \lambda^a  t
               {\bar t} \gamma_5  \lambda^a  b
        -    {\bar b}   \lambda^a  t
                {\bar t} \gamma_5  \lambda^a  b
        +       {\bar b} \gamma_5  \lambda^a  t
                     {\bar t}   \lambda^a  b
              \right] .
\end{equation}
Ignoring the bottom quark degree of freedom at this point,
because their Yukawa coupling is much
lower than that of the top mass, the usual bosonization procedure
\cite{bolo:egu} leads to
\begin{equation}
  {\cal L}_{int}^{tt}      =
       {f_t}\   {\bar t}
          \left( {\bar \sigma} + i {\bar \pi} \gamma_5 \right)
                  t
               +  {\chi^2\over 4}
       {\rm tr} \left( {\bar \sigma}^2 +     {\bar \pi}^2  \right)
\label{lagrange2}
\end{equation}
where $G/2=f_t^2/\chi^2$ and
${\bar \sigma}$ and ${\bar \pi}$ are real SU(3) matrix fields.
These are in the adjoint representation of $SU(3)_c$.
Integrating out the quarks gives the one loop effective action
\begin{equation}
   S_{eff}^{tt} =
       - {\rm Sp\ } \log \left(
         - i  \rlap{/}{\partial}  +   {f_t}
       ({\bar \sigma} + i {\bar \pi} \gamma_5)
          \right) +
        {\chi^2\over 4}  \int d^4x
      {\rm tr} \left( {\bar \sigma}^2 +     {\bar \pi}^2  \right)
\label{t19}    
\end{equation}
A stationary phase condition for the vacuum leads
to a gap equation of the spontaneously broken phase
with ${\bar \sigma}_v\ne 0$  and
\begin{equation}
 \chi^2 = 8 f_t^2 I_1(M_{con}) ,
\label{t20}
\end{equation}
where $M_{con}=f_t{\bar \sigma}_v$ is a diagonal 
constituent quark mass matrix for
the top. As we will see later, this is not 
the physical mass of the top
but rather the classical one which will be modified
in the presence of the non-homogeneous Dirac sea.
In eq.\ (\ref{t20})
\begin{equation}
   I_n(M) = {1 \over 16\pi^2} 
\int {du \over u^{3-n} }
         e^{-u M^2 } \phi(u),\ \ \ \ \phi(u)= \sum_{i=1}^n \theta
                (1- {1\over\Lambda_i^2} )
\label{prot} 
\end{equation}
are proper-time regularized \cite{bolo:ebre} Feynman integrals.
We gave in eq.\ (\ref{prot}) a rather 
general form of the cutoff function
as a sum over n step like cutoff functions \cite{bolo:ab4}.
However for our present purpose we will
take $n=1$, which is sufficient to render the model finite.
More cutoffs could be used to fix condensates 
or current masses \cite{bolo:ab4}.
In choosing the vacuum state ${\bar \sigma}_v\ne 0$ in
eq.\ (\ref{t20}) the original $SU(3)_R\otimes SU(3)_L$ symmetry is
spontaneously broken down to $SU(3)_V$.
As a result the pion-like fields ${\bar\pi}$ 
are the Goldstone bosons of
the spontaneously broken symmetry.
The usual wave-function renormalization condition $Z_\pi=1$
\cite{bolo:egu,bolo:mrg,bolo:klein} gives
\begin{equation}
	4  {M_{con}^2 } I_2(M_{con}) = f_\pi^2
\label{fpi}
\end{equation}
and determines the value of the cutoff $\Lambda$ (\ref{prot})
in terms of the constituent mass $M_{con}$. 
Thus for a given top-pion
decay constant $f_\pi$, the top-constituent 
mass $M_{\rm con}$ is the only
free parameter of the model. The top-pion decay constant
is of course not yet known but it is expected to be of the order of
the Fermi scale. We choose a central value of
\begin{equation}
  f_\pi=f_\pi^{tt} \simeq  30 {\rm GeV}
\end{equation}
which is of the order of $f_\pi^{tb}\simeq 50{\rm GeV}$
proposed by Hill \cite{bolo:hill} in some related technicolor model.
However this fixing is by no means unique 
and we also studied variations
of this central value in Tab. (\ref{tab2}) and  Tab. (\ref{tab3}).
However -- as we shall see later -- the anomalous 
magnetic moment will
turn out to be independent of $f_\pi$. 
%
%
\section{Solitons}
In order to describe a baryon number $\frac{1}{3}$ 
system from the effective action
eq.\ (\ref{t19}), we have to consider meson fields 
$\bar{\sigma}$ and $\bar{\pi}$
different from the vacuum configuration. 
Because we are interested in static
properties of the top, we restrict ourselves to 
time-independent meson
fields.

Considering first the SU(2) version of the model, 
one has to make the replacement
\begin{equation}
      {\bar \sigma} + i {\bar\pi} \gamma_5 \longrightarrow
      f_\pi  \xi =
      f_\pi e^{i \theta {\hat x} {\vec \tau} \gamma_5 } .
\end{equation}
This combines two things. First the usual 
hedgehog ansatz for the quark fields
allows only the isoscalar scalar and 
isovector pseudoscalar fields to have
non-trivial equation of motion. Second a non-linear constraint for
$\bar{\sigma}_0$ and $\bar{\pi}$ has to be used in order to get
stable solitonic solutions \cite{bolo:sieber}. So the meson field
configuration is given by
the chiral angle $\theta$, which can be obtained from the
effective action eq.\ (\ref{t19}) by solving the 
corresponding equations of motion
\cite{bolo:megrgo,bolo:rewu,bolo:cbk96}.

By defining a {\bf Giantspin} ${\vec P}={\vec j} + {\vec \lambda}$
analogous to the usual Grandspin
${\vec G}={\vec j} + {\vec \tau}$, the hamiltonian
\begin{equation}
    H = { \vec \alpha \cdot \vec \nabla \over i}  + \beta
    M_{con} \left( \cos \theta(r) + i\gamma_5 \sin \theta(r)
	\frac{\vec x}{r} \cdot {\vec \lambda} \right)
	\label{hamiltonian}
\end{equation}
commutes with Giantspin $P^2$, its z-component 
$P_z$ as well as the parity
$\Pi$ and these four quantities form a complete 
set of commuting operators.
The eigenstates $|n\rangle $ of $H$:
\begin{equation}
     H|n\rangle = E_n |n\rangle \qquad
     \langle \vec{r} \mid n \rangle = \phi_n(\vec{r})
     \label{eigenproblem}
\end{equation}
can be characterized by four quantum numbers 
$E_n, P, P_z$ and $(-)^l$,
where $P_z$ is degenerated. The eigenvalue problem 
eq.\ (\ref{eigenproblem}) will
now be solved numerically following \cite{bolo:kari} 
by putting the system in a large,
finite box and demanding appropriate boundary conditions 
for the radial part
of the eigenfunctions at the end of the box. So we 
obtain a discrete basis,
which belongs to a given giant spin $P$ and parity 
$\Pi$. With a numerical
cutoff the basis became finite, so that a numerical 
diagonalization is
possible. Note that this numerical cutoff has nothing 
to do with the physical
UV cutoff $\Lambda$ mentioned above.

The equations of motion for our system are given by 
stationary points of the
effective chiral action. These are solvable via a 
standard selfconsistent
procedure as it is known from Hartree-Fock 
calculations in nuclear
physics \cite{bolo:megrgo2,bolo:rewu}.
We start with a reasonably chosen profile 
$\theta (r)$, diagonalize the $H$ of eg.\ (\ref{hamiltonian}) 
as described above, and obtain eigenfunctions 
$\phi_n(\vec{r})$ and eigenvalues
$E_n$, which lead to a new profile through the 
equations of motion. This
selfconsistent procedure is iterated until a 
reasonable small difference in
the profiles is reached.

In general the selfconsistent procedure is performed 
for a given value of the
constituent top mass $M_{\rm con}$ (or equivalently 
the scaled cutoff
$\Lambda / M_{\rm con}$, see Tab. (\ref{tab2})). 
Furthermore explicit symmetry
breaking current masses as well as no symmetry 
breaking current mass have been
considered (compare Tab. (\ref{tab3}) and Tab. (\ref{tab4})). 
It turns out,that solitonic
solutions exists only if the scaled cutoff $\Lambda$ 
lies under a critical value
$\Lambda < \Lambda_{\rm cr} \approx 2$, which corresponds 
to lower bound of 
the constituent 
top mass of $M_{\rm cr} \approx 180$ GeV for $f_\pi = 30$ GeV. 
This critical 
bound is typical for localized
solutions of a system of coupled non linear equations 
\cite{bolo:tdlee}
(solitonic solutions).

In dependence of the radial size of the profile $\theta$, 
one finds bound 
orbitals both in the positive and negative spectrum of 
eigenstates. Due to the
spontaneous symmetry breaking there is a gap of size 
$2\cdot M_{\rm con}$
between these parts of the spectrum and the positive 
bound state with lowest
single particle energy $E_n$ and quantum numbers 
$P^+ =0^+$ will be called
valence orbit $E_{\rm val}$. Its energy decreases 
with increasing  profile
size, switches sign and gets finally part of the 
negative spectrum
\cite{bolo:megrgo2}. This
valence level, coming from the positive continuum, 
gets bound and localized by
interacting with the negative continuum, which gives 
us the solitonic solution of the system.

An ansatz for the SU(3) fields is given by using the 
trivial embedding
\cite{bolo:wit83b,bolo:kaplan} of the subgroup of 
SU(2) into SU(3).
Then the unitary field $\xi$ performs to
\begin{equation}
      \xi = \left(\begin{array}{cc}
        e^{ i \theta(r) {\hat x} {\vec\lambda} \gamma_5 } & 0 \\
        0 & 1   \end{array} \right) ,
\end{equation}
This embedding made by
Witten for the flavor SU(3) group is distinguished by giving the
correct selection rules for the low lying multiplets with definite
color and spin.

The hedgehog mean field solution of the classical 
equations of motion breaks
the rotational symmetry of the full theory, i.e. 
the eigenstates
$|n\rangle $ of $H$
do not carry good spin quantum number. To restore 
this symmetry, we couple
the corresponding spin expectation value to the 
effective chiral action, which
turns out to be equivalent to consider the soliton 
in a rotating system, called
cranking method\cite{bolo:dipepo,bolo:goetal}.
The main idea is to perform an adiabatic rotation of the
hedgehog fields with angular velocity $\vec{\Omega}$ 
and treat the problem
perturbatively in $\Omega$. 
In this expansion there occur rotational 
terms subleading in $\Omega$.
They are not necessarily small due to their 
-- long time overlooked --
origin\cite{bolo:ab9,bolo:ab10,bolo:allstars}: 
The model is formulated
in terms of a 
fermion determinant (\ref{t19}), so that the 
operators, which occur in
the path integral formulation for matrix 
elements have to be explicit
time-ordered. Supplementary the cranking 
velocities of the
semiclassical quantization turn out to be 
collective operators, which
do not commute with the rotation matrix itself.

The cranking procedure gives us the moment 
of inertia tensor
\begin{equation}
I_{AB} = \frac{1}{4} \int \frac{d\omega}{2\pi}
{\rm Tr}_{\gamma\lambda_c} 
{\frac{1}{i\omega + H}\lambda_A\frac{1}{i\omega + H}\lambda_B} .
\end{equation}
Because of the embedding we have
\begin{equation}
I_{AB} = \left\{ \begin{array}{cl}
	 I_1 \delta_{AB} & \mbox{for} A,B = 1,2,3 \\
	 I_2 \delta_{AB} & \mbox{for} A,B = 4,5,6,7 \\
	 0 & \mbox{for} A,b = 8 \\
		\end{array} \right.
\end{equation}
and by defining right generators $R_a$ \cite{bolo:mnp,bolo:pras2}
\begin{eqnarray}
   R_a  = \left\{ \begin{array}{cl}
   I_1 \Omega_a & a=1,2,3 \\
   I_2 \Omega_a & a=4,5,6,7 \\
   {1\over 2  \sqrt{3} } := {\sqrt{3}\over 2} Y_R & a=8 \\
       \end{array} \right.	
\end{eqnarray}
we can see that the right hypercharge is restricted to $1/3$.
This corresponds to
multiplet representations with triality 1.  The 
lowest SU(3)
representation with unit triality is therefore 
the triplet {\bf [3]} with
spin $1/2$. The next one is the color antisextett 
with spin $1/2$,
which can be considered as some $t({\bar t}t)$ 
excitation.
Because we have $N_c=1$, the baryon number and
therefore the right hypercharge $Y_R$ is now $1/3$ 
and the lowest
possible representation therefore is the triplet 
{\bf [3]}, the
fundamental representation of SU(3). By expressing 
the spin expectation
value and the rotational contribution of $S_{\rm eff} 
[\Omega]$ in terms
of the SU(3) rotation matrix we can perform the 
quantization in the
standard way by substituting the coordinates and 
conjugate momenta
through the operators which should fulfill the 
usual commutator rules.
After subtracting the spurious zero mode contributions 
this leads to
\begin{eqnarray}
 M = M_{cl}    + { 1 \over 2 I_2 } C_2(SU(3))
 + \left( { 1 \over 2 I_1 } - { 1 \over 2 I_2 } \right)
 C_2(SU(2)) - { 3 \over 8 I_1 } - {7 \over 24 I_2 } ,
 \label{mtop}
\end{eqnarray}
where $C_2(SU(2)_R)=j(j+1)=3/4$ and
$C_2(SU(3)_{R/L})=1/3(p^2+q^2+3(p+q)+pq)=4/3$ for the 
color triplet
{\bf [3]} and spin $1/2$ representation (see e.g.
\cite{bolo:kaplan,bolo:zhang1}).  So in the lowest 
representation
$(3,\frac{1}{2})$ with $(p=1,q=0)$ the physical top mass 
$M_{top}$ is simply
given by
\begin{equation}
	M _{top} =  M^{(3,\frac{1}{2})}  = M_{cl} .
\end{equation}
According to our
numerical results, to be presented in sect. \ref{sec6}, 
the next
higher representation  $({\bar 6},\frac{1}{2})$ with  $(p=0,q=2)$ 
is shifted
by\footnote{Note that this shift differs from eq. (4.29) in
\cite{bolo:kaplan}.}
\begin{equation}
	M^{({\bar 6},\frac{1}{2})} - M^{(3,\frac{1}{2})}
	= { 1 \over I_2 }  \simeq 380 {\rm GeV}
\end{equation}
for $M_{con}=190{\rm GeV}$,
$f_\pi=30{\rm GeV}$, whereas   $(15,\frac{1}{2})$ 
in the $(p=2,q=1)$ multiplet
gives
\begin{equation}
	M^{(15,\frac{1}{2})} - M^{(3,\frac{1}{2})}
        = { 2 \over I_2 }  \simeq 760 {\rm GeV} .
\end{equation}
The next higher spin multiplet
$(15,{3\over 2})$  gives
\begin{equation}
  	M^{(15,{3\over 2})} - M^{(3,\frac{1}{2})}
        = { 3 \over 2I_1 } + { 1 \over 2I_2 } \simeq 460 {\rm GeV}
\end{equation}
which leads to the remarkable fact, that the  spin
$3/2$ multiplet is below a spin $1/2$ multiplet.

The moments of inertia in (\ref{mtop})
are decomposed into {\it valence} and {\it sea} parts
$I_k=I_k^{val}+I_k^{sea}$, $k=1,2$, according to
\begin{eqnarray}
	I_1^{val}  &=& {1\over 6} \sum_{i=1}^3 \sum_{n\neq val}
             {   \langle val\mid\lambda_i\mid n \rangle
                \langle n\mid\lambda_i\mid val \rangle
               \over  E_{val} - E_n },\ \ \ i=1,2,3
                \nonumber \\
       I_2^{val}  &=& {1\over 8} \sum_{a=4}^7 \sum_{n\neq val}
           {     \langle val \mid\lambda_a\mid n \rangle
                \langle n\mid\lambda_a\mid val \rangle
             \over  E_{val} - E_n },\ \ \ a=4,5,6,7
\end{eqnarray}
And
\begin{eqnarray}
I_1^{sea}  &=& {1\over 12} \sum_{i=1}^3 \sum_{m,n}
                \langle m\mid\lambda_i\mid n \rangle
                \langle n\mid\lambda_i\mid m \rangle
              {\cal R}_I(E_n,E_m),\ \ \ i=1,2,3     \nonumber \\
I_2^{sea}  &=& {1\over 16} \sum_{a=4}^7 \sum_{m,n}
                \langle m\mid\lambda_a\mid n \rangle
                \langle n\mid\lambda_a\mid m \rangle
              {\cal R}_I(E_n,E_m),\ \ \ a=4,5,6,7
\end{eqnarray}
where ${\cal R}_I(E_n,E_m)$ is the usual 
regularization function for
the sea part of moment of inertia \cite{bolo:ab4}:
\begin{eqnarray}
    {\cal R}_I(E_n,E_m)  &=& -\frac{1}{2\pi} \int_0^\infty
	\frac{du}{\sqrt{u}} \sum_{i=1}^n \theta\left( u - {1\over
   \Lambda_i^2}  \right) \frac{1}{E_m + E_n} \times \nonumber \\
	&& \left[ E_n \exp (-u E_n^2) + E_m \exp (-u E_m^2) +
		\frac{\exp (-u E_n^2)-\exp (-u E_m^2) }
			{ u (E_n - E_m) } \right] .	
\end{eqnarray}
%
%
\section{Chromomagnetic moments}
The color octet of gauge fields $G_\mu^a(x)$ couple in the QCD
Lagrangian to the top quarks
$    {\cal L}_{ttG} = g\  j_\mu^a(x)
       \  G_\mu^a(x) $
where
$j_\mu^a(x)={\bar t}(x) \gamma_\mu\lambda^a\frac{1}{2} t(x)$ 
is the conserved
color octet vector current. From this we can define the
{\bf chromomagnetic moment} of the top as
\begin{equation}
	\mu_z^c  = {g\over 2 M_{top} }  \int d^3x \left( {\vec x}
		\times {\vec j}^c  \right)_z = : { g \over 2M_{top}}
			G_M^c(q^2=0)
\end{equation}
where $g$ is the gluon coupling constant, related to $\alpha_s$
in QCD. The general form of the matrix element of the 
vector current for a
localized top quark state can be described by 
{\bf color formfactors}
$F_1$ and $F_2$ as:
\begin{equation}
	\sqrt{ p_0 p_0' \over M_{top}^2}
 \langle t(p,c,s)\mid j_\mu^a(0) \mid t(p',c,s) \rangle
 = {\bar t}_{c,s} (p) \left[
 F_1(q^2) \gamma_\mu + i{ \sigma_{\mu\nu}q_\nu\over 2 M_{top}}
 F_2(q^2) \right]  \lambda^c \frac{1}{2} \     t_{c,s} (p') .
\end{equation}
From the color formfactors one can calculate the chromoelectric
$G_E^c(q^2)$ and chromomagnetic $G_M^c(q^2)$ 
formfactors according to
\begin{eqnarray}
 G_E^c(q^2) &=& G_E(q^2) L_c 
= \left[F_1(q^2)+{q^2\over 4 M_{top}^2 }
 F_2(q^2)   \right]    L_c
 \nonumber \\  
G_M^c(q^2) &=& G_M(q^2) L_c  = \left[F_1(q^2)+ F_2(q^2)
 \right] L_c
\end{eqnarray}
where $L_c$ is eigenvalue of the left color SU(3) 
generator. Now it can
be shown that $F_1(0)$ is related to the normalized 
color charge  and
therefore $F_1(0)=1$. Then the {\it anomalous} 
chromomagnetic moment $\kappa=F_2(0)$ follows as
\begin{equation}
	\kappa=F_2(0)= G_M(0) - G_E(0) = G_M(0) - 1 .
\end{equation}
Remember that for a point like particle $F_1=1$ 
and $F_2=0$.
Similarly in low energy QCD regime we have for 
the proton
$F_1=1$ and $F_2=1.79$. Therefore the prediction 
of a non-zero
$\kappa=F_2(0)$ is a result of the color form factor
of the top and is absent for point-like 
quarks\footnote{Provided that
non-universal interactions are neglected
\cite{bolo:zhanglee,bolo:whiszhang}.}.
However the reverse is not true. We will 
show that $\kappa=0$ and
$R\ne 0$ is consistent. Actually we show 
that $\kappa+1$ is
proportional to $M_{top}\cdot R$ in contrast to $\kappa$ 
itself in Ref. \cite{bolo:brosch}.

Similar to the axial current formfactor 
in Ref. \cite{bolo:ab9}
within a flavor SU(3) model
the chromomagnetic formfactor at $q^2=0$  can be written as
\begin{equation}
G_M^c(q^2=0) =
  \left( A + {B\over I_1} +  {C\over I_2} \right)
    D_{ci}
    -   {D\over I_2}
     d_{ibb} D_{cb} R_b  - {E\over I_1}
     D_{c8} R_i,\ \ i=3
\end{equation}
where the coefficients  $A,B,C,D,E$ are given by
\begin{eqnarray}
A  &=& { M_{top} \over 6}  \sum_{n}
  \langle n \mid  \gamma_0 \gamma_i 
   \lambda_3  {\hat x}_j \mid n\rangle      \epsilon_{i3j}
   {\cal R}_\Sigma(E_n)
   \nonumber \\
   B  &=&  { M_{top} \over 12 } \sum_{m,n} \langle n\mid
	\gamma_0 \gamma_i \lambda_k {\hat x}_j
    \mid m\rangle  \langle m\mid \lambda_l \mid  n\rangle
	\epsilon_{ij3} \epsilon_{kl3}
              {\cal R_Q}(E_n,E_m)
        \nonumber \\
      C  &=&
       { M_{top} \over 12 } \sum_{m,n} \langle n\mid  \gamma_0
	\gamma_i \lambda_a  {\hat x}_j
            \mid m\rangle  \langle m\mid  \lambda_b \mid n\rangle 
	\epsilon_{ij3} \left( f_{3ab} - \epsilon_{3ab} \right)
           {\cal R_Q}(E_n,E_m)
           \nonumber \\
      D  &=&
           { M_{top} \over 12 } \sum_{m,n} \langle n\mid  
	\gamma_0 \gamma_i \lambda_a {\hat x}_j
    \mid m\rangle  \langle m\mid \lambda_a   \mid  n\rangle \
	\epsilon_{ij3} \  d_{3aa}
            {\cal R_M}(E_n,E_m)
           \nonumber \\
      E  &=&
  { M_{top} \over 12 } \sum_{m,n} \langle n\mid  \gamma_0 \gamma_i
	\lambda_a  {\hat x}_j
     \mid m\rangle  \langle m\mid \lambda_b   \mid  n\rangle \
	\epsilon_{ij3}\   \delta_{a8} \  \delta_{b3}
          {\cal R_M}(E_n,E_m)
           \nonumber \\
\end{eqnarray}
The regularization functions are defined by \cite{bolo:ab10}:
\begin{eqnarray}
{\cal R}_\Sigma (E_n) &=& \frac{1}{\sqrt{\pi}}
\int_0^\infty
\frac{du}{\sqrt{u}}
\exp (-u) \sum_{i=1}^n c_i
\theta
\left( \frac{u}{E_n^2}-{1\over \Lambda_i^2}  \right) \nonumber \\
{\cal R_Q} (E_n,E_m) &=& \int_0^1 \frac{d\alpha}{2\pi}
\frac{\alpha E_n - (1-\alpha) E_m}{\sqrt{\alpha(1-\alpha)}}
 \sum_{i=1}^n
\  c_i
\cdot \frac{\exp \left( -\frac{\alpha E_n^2 + (1-\alpha) E_m^2}
{\Lambda_i^2}\right)}
{\alpha E_n^2 + (1-\alpha) E_m^2} \nonumber \\
{\cal R_M} (E_n,E_m) &=& \frac{1}{2}
\frac{{\rm sign} (E_n - \mu) - {\rm sign} (E_m - \mu) }{E_n - E_m} .
\end{eqnarray}
The collective integrals reduce to \cite{bolo:keaton}
\begin{equation}   \langle t(c,s)\mid D_{ai}   \mid t(c,s) \rangle
        = \langle D_{ai}  \rangle_c
         = -{3\over 4}  S_i  L_a
\end{equation}
where $S_m$ is the eigenvalue of the SU(2) spin generator
and $L_n$ is the eigenvalue of the left SU(3) color generators.
Similarly
\begin{equation}
       \langle t(c,s)\mid d_{ibb} D_{ab} R_b   \mid t(c,s) \rangle
      = \langle d_{ibb} D_{ab} R_b \rangle_c
         = - {3\over 8}  \ \   S_i  L_a
\end{equation}
\begin{equation}   
\langle t(c,s)\mid D_{a8} R_i  \mid t(c,s) \rangle
      = \langle  D_{a8} R_i  \rangle
         = { {\sqrt 3}\over  8}\ \   S_i  L_a
\end{equation}
%
%
\section{Numerical results} \label{sec5}
As mentioned already, for a given top-pion decay 
constant $f_\pi$ the
constituent top mass $M_{\rm con}$ is the only 
free parameter of the
model in the mesonic as well as in the solitonic sector. 
The $f_\pi$ is
chosen between 25 GeV and 50 GeV according to \cite{bolo:hill}. 
For the actual
calculation we use $f_\pi = 30$ GeV. For this $f_\pi$ the 
constituent top mass
is chosen to $M_{\rm con} = 190$ GeV since then a soliton 
top mass around $M_{top} = 190$ GeV is achieved. Here we make the
assumption that the constituent top mass 
$M_{}con$ and the soliton top
mass, $M_{top}$, should not differ to much. Otherwise the model and
the concept of a soliton are not well defined. The corresponding
scaled cutoff is then determined at
$\Lambda / M_{\rm con} = 1.9$. Keeping $\Lambda / M_{\rm con}$ 
fixed, we varied
$f_\pi$ between 25 GeV and 50 GeV with, however, 
negligible effect on the
chromomagnetic anomalous moment $\kappa$ (Tab. (\ref{tab3})). 
Hence all further
and detailed calculations have been done with $f_\pi = 30$ 
GeV and no further
adjusting of parameters has been performed. The fact that a variation
of $f_\pi$ has only a small effect on the anomalous chromomagnetic
moment $\kappa$ can be explained in the following way:
Due to the dimension MeV of $f_\pi$, the 
influence of this constant on 
observables of our model can be estimated by the dimension of the
observables themselves. For example the mean squared radius scales
like $1/({\rm MeV})^2$, so it decreases with $1/f_\pi^2$ if $f_\pi$
increases. The anomalous chromomagnetic moment itself is a
dimensionless quantity, because its given 
by the chromomagnetic moment
(dimension $1/{\rm MeV}$) scaled by the top quark mass $M_{top}$. So
one should expect that this anomalous moment does not depend
significant on $f_\pi$, which can be seen in Tab. (\ref{tab3}).

The final results for $\kappa$, including rotational 
corrections to order
${\cal O }(\Omega^1)$, are presented in Tab. (\ref{tab3}) 
and Fig. (2).
Apparently the chromomagnetic anomalous moment $\kappa$ 
lies between 0.2
and -0.2 for a range of constituent top quark masses 
$M_{\rm con}$, which
agree roughly with the solitonic mass and are less 
than 350 GeV. It is
interesting to investigate the effect of an explicit 
symmetry breaking by
giving the current top quark a finite mass. The resulting 
effect on $\kappa$
with a current mass of $m = 60$ GeV is shown for 
illustration in Fig. (2)
as well.

Animated by Brodsky and Schlumpf \cite{bolo:brosch}, 
we plotted the anomalous
magnetic moment against the product of the top soliton 
mass and the radius
$R = \sqrt{<r^2>}$ (see Fig.\ (3)). Surprisingly the 
result shows that
$(\kappa + 1)$ is proportional to $M_{top}\cdot R$, which is 
supported by a linear
fit of our data, shown in Fig.\ (3) too.

Finally we used a relation between the top production 
rate and $\kappa$ given
by Atwood, Kagan and Rizzo \cite{bolo:akr} to combine 
it with our relation
between $\kappa$ and $M_{\rm con}$. So we are able to 
give the top production
rate as a function of the top constituent mass $M_{\rm con}$ 
(Fig.\ (4)),
the only free parameter of our model.
%
%
\section{Summary} \label{sec6}
We have shown that at some high energy scale $\Lambda$ 
the spontaneous
breaking of SU(3) color not only explains the mass of 
the top quark in the
electroweak theory but also explains an anomalous 
chromomagnetic moment
of the top quark.

To this aim we use an effective color $SU (3)\otimes SU(3)$ 
invariant
Lagrangian that describes the mass of the top via 
dynamical symmetry
breaking. Soliton solutions are obtained by trivially 
embedding an
SU(2) hedgehog Ansatz for the color pion octet into 
SU(3) \cite{bolo:wit83b}
and minimizing the classical energy functional. This 
gives a classical
mean field solution in terms of an chiral angle.
The quantization of the soliton is done by 
semiclassical quantization
of the rotational zero modes of the time independent 
classical solution.
Imposing canonical commutation rules for the 
collective coordinates,
generators of SU(3) color and SU(2) spin can 
be identified.
Due to the trivial embedding  the $(3,1/2)$ 
triplet with spin 1/2
appears as lowest possible representation. 
Higher spin and color
states turn out to be shifted by roughly two 
times the top mass.

The chromomagnetic moment turns out to be positive 
and of the order
$-0.16 \leq \kappa \leq +0.13$  provided the constituent 
top quark mass is
of the order of the soliton mass.
Therefore it has the correct magnitude to explain
the recent measured top production rate at the 
Tevatron \cite{bolo:akr}.
To this aim we calculated the chromomagnetic 
moment in a solitonic background
field for a  pseudoscalar octet of Goldstone 
pions  up to the
linear order in the rotational velocity of the 
semiclassical quantization.
It turned out that the recently found  corrections 
in the linear order from
explicit time-ordering of collective operators  
\cite{bolo:ab9,bolo:allstars}
give important contribution to the anomalous magnetic moment.

So we used a model, which is able to describe a 
top-quark via solitonic
solutions in a quite reasonable way. If the cross 
section of top quark
production was eventually found to agree with the 
SM expectations, the
estimated bounds on $\kappa$ -- obtainable on the 
Tevatron -- are given
by $-0.14 \leq \kappa \leq 0.15$ for 
luminosity ${\cal L} = 100\,{\rm
pb}^{-1}$ up to $-0.08 \leq \kappa \leq 0.11$ 
for ${\cal L} = 1000\,{\rm
pb}^{-1}$, if all errors on the experimental and theoretical side 
are taken into account \cite{bolo:akr}. Our presented model gives 
an anomalous magnetic moment, which lies in the estimated bounds and
also gives the new D0 and CDF results for 
the top cross section with a
top mass in the physical magnitude.

On the other hand, if the result of D0 Collaboration 
\cite{bolo:d0}  are
correct, -- and the new CDF measurements support 
this -- there is no
direct need to have an anomalous 
chromomagnetic moment far away from zero.
However a high statistic measurement will be necessary to
determine values of kappa [-0.1,0.1], which would be naturally
supported in this soliton picture. But we should stress that even
in the limit
of vanishing kappa, our soliton model provides a 
picture for the top
quark as a valence quark bound in a cloud of color 
octet Goldstone
bosons. As a consequence it can have  a non-vanishing radius
of the order ${\cal O}(1am)$, contradicting the
former picture of having kappa proportional to
$M_{top}\cdot\ R$ \cite{bolo:brosch,bolo:brodre}.

Altogether our model may be a good instrument to 
investigate the new physics
of the top quark in a quite simple, but effective 
way. The next high
statistic measurements of D0 and CDF
of the near future will decide,  if this model can 
tell us more about the top.
%
%
\acknowledgements
We would like to thank Michal Praszalowicz (Krakow) for useful
discussions concerning the quantization 
for SU(3). Also we would like
to thank Jack Smith (SUNY Stony Brook) for bringing the recent
D0 measurements and SM implications to our attention.
One of us (AB) would like to thank the {\it Alexander von Humboldt
Foundation} for a Feodor Lynen grant.
\newpage
\appendix
\section{SU(3) matrix elements}

Wave functions are given by
\begin{equation}
  \Psi_{(Y,T,T_3)(Y_R,J,J_3)}^{(3)}(A) =
     \sqrt{3} (-)^{Y_R/2+J_3} D_{(Y,T,T_3)(-Y_R,J,-J_3)}^{(3)*}(A)
\end{equation}
with $Y_R=-{N_c\over 3}=-{1\over 3}$ for convenience and
$J=\frac{1}{2}$ and $J_3=\frac{1}{2}$ for the spin $1/2$ triplet.
The $D_{(Y,T,T_3)(-Y_R,J,-J_3)}^{(3)*}(A)$
are the usual Wigner wave functions  \cite{bolo:DS}.
Then
\begin{equation}
  \Psi_{(Y,T,T_3)(-1/3,1/2,1/2)}^{(3)}(A) =
     \sqrt{3} (-)^{1/3} D_{(Y,T,T_3)(1/3,1/2,-1/2)}^{(3)*}(A) ,
\end{equation}
where possible combinations for the color indices in {\bf [3]} are
\begin{equation}
  (Y,T,T_3) =  \left( 
\begin{array}{ccc} 1/3 & 1/2 & 1/2 \nonumber \\
                                 1/3 & 1/2 & -1/2 \nonumber \\
                                 -2/3 & 0   & 0  \end{array} \right)
                =
                \left( \begin{array}{c} b \nonumber \\ 
g \nonumber \\ r    \end{array} \right)
             = color .
\end{equation}
Using that ${\bf[8]}\otimes{\bf [{\bar 3}]}={\bf [{\bar 3}]}
+{\bf [6]}+{\bf [15]}$ we find
\begin{eqnarray}  
& & \int d\xi_A  D_{(Y,T,T_3)(1/3,1/2,-1/2)}^{(3)}(A)
       D_{ab}^{(8)}(A)
       D_{(Y,T,T_3)(1/3,1/2,-1/2)}^{(3)*}(A)
      =  \nonumber \\ & &
           \left( \begin{array}{ccc}
        8  &  {\bar 3}  & {\bar  3}  \nonumber \\
        a  &  color  & color    \end{array} \right)
         \left( \begin{array}{ccc}
        8  &  {\bar 3}  & {\bar  3}  \nonumber \\
        b  & -{1\over 3},\frac{1}{2},\frac{1}{2}  
&-{1\over 3},\frac{1}{2},\frac{1}{2}    \end{array} \right) ,
\end{eqnarray}
where we used the relation between complex conjugation and
anti-multiplets:
\begin{equation}
        D_{(L)(R)}^{(3)*}(A)
        =    D_{(-L)(-R)}^{({\bar 3})}(A)
                  (-)^{Q(L)-Q(R)}
\end{equation}
with the charge of the states $(-R)=(-Y_R,J,-J_3)$ and
$Q(R)=J_3+Y_R/2$. Therefore in our case we have to consider
\begin{eqnarray}
	\langle t(c,s) \mid  D_{33}^{(8)}(A)
            \mid     t(c,s)  \rangle  &=&
           \left( \begin{array}{ccc}
        8  &  {\bar 3}  & {\bar  3}  \nonumber \\
        010  &  -color  & -color    \end{array} \right)
         \left( \begin{array}{ccc}
        8  &  {\bar 3}  & {\bar  3}  \nonumber \\
   010   &  -{1\over 3},\frac{1}{2},\frac{1}{2}  
& -{1\over 3},\frac{1}{2},\frac{1}{2}    \end{array} \right)
          \nonumber \\
        \langle t(c,s) \mid  D_{38}^{(8)}(A)
            \mid     t(c,s)  \rangle  &=&
         \left( \begin{array}{ccc}
       8  &  {\bar 3}  & {\bar  3}  \nonumber \\
       010  &  -color  & -color    \end{array} \right)
        \left( \begin{array}{ccc}
       8  &  {\bar 3}  & {\bar  3}  \nonumber \\
  000   &  -{1\over 3},\frac{1}{2},\frac{1}{2}  
& -{1\over 3},\frac{1}{2},\frac{1}{2}    \end{array} \right)
           \nonumber \\
        \langle t(c,s) \mid  d_{3bb} D_{3b}^{(8)}(A)
            \mid     t(c,s')  \rangle  &=&
          \left( \begin{array}{ccc}
        8  &  {\bar 3}  & {\bar  3}  \nonumber \\
        010  &  -color  & -color    \end{array} \right)
         \left( \begin{array}{ccc}
        8  &  {\bar 3}  & {\bar  3}  \nonumber \\
        b  &  -spin'    &  -spin   \end{array} \right) d_{3bb}
\end{eqnarray}
and the relation between complex conjugation and anti-multiplets:
\begin{equation}
	D_{(L)(R)}^{(3)*}(A)
        =    D_{(-L)(-R)}^{({\bar 3})}(A)
                  (-)^{Q(L)-Q(R)} .
\end{equation}
Noting that \cite{bolo:currents,bolo:toyota}
\begin{equation}
	\begin{array}{rrrrr}
  V_-\mid   {\bar 1} \rangle   &=& R_{4+i5} \mid {\bar 1} \rangle
   &=&  \mid {\bar 3}
  \rangle  \nonumber \\
  V_+\mid   {\bar 3} \rangle &=& R_{4-i5} \mid {\bar 3} \rangle &=&
      \mid {\bar 1}
   \rangle   \nonumber \\
   U_-\mid   {\bar 1} \rangle  &=&  R_{6+i7} \mid {\bar 1} \rangle
  &=&
     \mid {\bar 2} \rangle    \nonumber \\
    U_+\mid   {\bar 2} \rangle &=&  R_{6-i7} \mid {\bar 2} \rangle
   &=&
   -   \mid {\bar 1}
   \rangle   \nonumber \\
  I_-\mid   {\bar 2} \rangle  &=&  R_{1+i2}  \mid {\bar 2} \rangle
 &=&    \mid {\bar 3}
  \rangle     \nonumber \\
    I_+ \mid   {\bar 3} \rangle &=&  R_{1-i2} \mid {\bar 3} \rangle
  &=&    \mid {\bar 2}
  \rangle    \nonumber \\
  I_3 \mid   {\bar 3} \rangle &=&  R_{3} \mid {\bar 3} \rangle &=&
   - 1/2    \mid {\bar 3}   \rangle  \nonumber \\
  I_3 \mid   {\bar 2} \rangle &=&  R_{3} \mid {\bar 2} \rangle &=&
    + 1/2      \mid {\bar 2}
  \rangle
  \end{array}
\end{equation}
and rewriting  $d_{3bb} D_{3b}^{(8)} R_b$ as
\begin{equation}   
	d_{3bb} D_{3b}^{(8)} R_b
       =  {\sqrt{2}\over 4} \left[
       - D_{3p} V_+   +  D_{3\Xi^-} V_-
       - D_{3n} U_+   -  D_{3\Xi^0} U_- \right]
\end{equation}
gives finally, with the SU(3) Clebsch-Gordon Coefficients
\cite{bolo:kaeding,bolo:kaeding2}.
\begin{eqnarray}
	\langle D_{33}^{(8)} \rangle &=&  -{3\over 4} S_3 F_3  \nonumber \\
 \langle D_{38}^{(8)} R_3 \rangle &=&  {\sqrt{3} \over 8} S_3 F_3
 \nonumber \\
 \langle d_{3bb} D_{3b}^{(8)} R_b \rangle &=&  -{3\over 8} S_3 F_3
\end{eqnarray}
These matrix elements are sufficient to calculate the magnetic moment
up to the linear order of the rotational frequency.
\newpage
\begin{table}
\label{tab1}
\caption{SU(3) Clebsch Gordon Coefficients. Particle quantum numbers
can be read of from Fig. 1.}
\begin{center}
\begin{tabular}{cccc}
${\bf [8]}$ &  ${\bf [{\bar 3}]}$  & ${\bf [{\bar 3}]}$   &   \\
\hline
$\Sigma^0$  &  ${\bar 1}$  &  ${\bar 1}$  & 0   \\
$\Sigma^0$  &  ${\bar 2}$  &  ${\bar 2}$  & $-\sqrt{3\over 16}$ \\
$\Sigma^0$  &  ${\bar 3}$  &  ${\bar 3}$  & $+\sqrt{3\over 16}$ \\
$\Sigma^+$  &  ${\bar 3}$  &  ${\bar 2}$  & $ \sqrt{3\over  8}$ \\
$\Sigma^-$  &  ${\bar 2}$  &  ${\bar 3}$  & $-\sqrt{3\over  8}$ \\
$\Lambda $  &  ${\bar 1}$  &  ${\bar 1}$  & $\sqrt{1\over 4}$  \\
$\Lambda $  &  ${\bar 2}$  &  ${\bar 2}$  & $-\sqrt{1\over 16}$  \\
$\Lambda $  &  ${\bar 3}$  &  ${\bar 3}$  & $-\sqrt{1\over 16}$  \\
$\Xi^0   $  &  ${\bar 1}$  &  ${\bar 2}$  & $ \sqrt{3\over  8}$  \\
$\Xi^-   $  &  ${\bar 1}$  &  ${\bar 3}$  & $ \sqrt{3\over  8}$  \\
$p       $  &  ${\bar 3}$  &  ${\bar 1}$  & $ \sqrt{3\over  8}$  \\
$n       $  &  ${\bar 2}$  &  ${\bar 1}$  & $-\sqrt{3\over  8}$  \\
\end{tabular}
\end{center}
\end{table}
%
\begin{table}
\caption{The cutoff $\Lambda$, the scaled cutoff 
$\Lambda/M_{con}$, the
Goldstone decay constant
$f_\pi$ and the top condensate as a function of 
the top constituent
quark mass $M_{con}$.}
\label{tab2}
\begin{center}
\begin{tabular}{ccccc}
       $M_{con}$ [GeV]   &  $f_\pi$ [GeV]
         &  $\Lambda$ [GeV]         &     $\Lambda/M$  &
      $-\langle{\bar t}_a t_a\rangle^{1/3}$ [GeV]   \\
\hline
   185.00   & 30.00  & 368.54   &  1.9921 &    86.9087      \\
   190.00   & 30.00  &   365.49  & 1.9236 &    86.1445     \\
   200.00   &  30.00  &  360.87 &  1.8043 &    84.8325      \\
   250.00   &  30.00  &  356.09 &  1.4244 &    80.9788      \\
   350.00   &  30.00  &  382.30 &  1.0923 &    78.8805      \\
   450.00   &  30.00  &  422.20 &  0.9382 &    79.0976      \\
   550.00   &  30.00  &  466.00 &  0.8473 &    79.5127      \\
\hline
   158.30   &  25.0  &  304.59 &  1.9241  &    71.7927    \\
   190.00   &  30.0  &  365.49 &  1.9236  &   86.1455     \\
   316.70   &  50.0  &  609.12 &  1.9233  & 143.5710      \\
\end{tabular}
\end{center}
\end{table}
%
\begin{table}
\caption{The energy of the lowest lying top quark multiplet
$M_{top}$, the corresponding Goldstone decay constant
$f_\pi$, the moments of
inertia $I_1$ and $I_2$, as well as the anomalous
chromomagnetic moment $\kappa$ as a function of the top constituent
quark mass $M_{con}$.}
\label{tab3}
\begin{center}
\begin{tabular}{cccccccc}
   $M_{con}$ [GeV]  &  $M_{top}$ [GeV]   &
                       $f_\pi$ [GeV]
         & $I_1$ [am]  &  $I_2$ [am]
          &             $\kappa$   & $<r^2> [\mbox{am}^2]$ \\
\hline
185 & 224.5(140.7) & 30 &  1.269  &  .592    &   0.1309 & 2.8206 \\
190 & 225.1(132.6) & 30 &  1.114  &  .516    &   0.0923 & 2.4354 \\
200 & 225.8(122.9) & 30 &  .954   &  .440    &   0.0289 & 2.0734 \\
250 & 224.3( 91.3) & 30 &  .651   &  .304    &  -0.1625 & 1.4183 \\
350 & 215.7( 35.0) & 30 &  .458   &  .219    &  -0.3643 & 0.9858 \\
450 & 204.3(-22.0) & 30 &  .311   &  .147    &  -0.4625 & 0.7259 \\
550 & 195.8(-82.8) & 30 &  .291   &  .132    &  -0.5527 & 0.6561 \\
650 & 186.5(-145.3)& 30 &  .246   &  .111    &  -0.6083 & 0.5670 \\
\hline
158.3 & 187.6(110.6) & 25 & 1.336  &  .619    & 0.0915 & 3.5113 \\
190.0 & 225.1(132.6) & 30 & 1.114  &  .516    & 0.0923 & 2.4354 \\
316.7 & 375.4(220.3) & 50 &  .694  &  .319    & 0.1084 & 0.8712 \\
\end{tabular}
\end{center}
\end{table}
\begin{table}
\label{tab4}
\caption{Same as Tab. III but with an explicit symmetry breaking
         current quark mass of 60 GeV.}
\begin{center}
\begin{tabular}{cccccccc}
   $M_{con}$ [GeV]  &  $M_{top}$ [GeV]   &
                       $f_\pi$ [GeV]
         & $I_1$ [am]  &  $I_2$ [am]
          &             $\kappa$   & $<r^2> [\mbox{am}^2]$ \\
\hline
225 & 263.1(213.1) & 30 & 2.3516 & 1.2034 & 0.1566 & 4.2064 \\
230 & 264.7(194.3) & 30 & 1.2967 & 0.5597 & 0.0962 & 2.2825 \\
235 & 265.8(189.5) & 30 & 1.1169 & 0.4734 & 0.0661 & 1.9879 \\
240 & 266.6(185.8) & 30 & 1.0006 & 0.4206 & 0.0400 & 1.8029 \\
250 & 267.7(179.6) & 30 & 0.8468 & 0.3543 &-0.0055 & 1.5644 \\
\end{tabular}
\end{center}
\end{table}
\newpage

\end{document}